\documentclass[twocolumn]{aastex62}

\usepackage[T1]{fontenc}
\usepackage{amsmath}
\usepackage{natbib}
\usepackage{url}
\usepackage{graphicx}
\usepackage{xspace}
\usepackage{booktabs}

\graphicspath{{Figures/}}

\newcommand{\prot}{P_{\rm rot}}
\newcommand{\teff}{T_{\rm eff}}
\newcommand{\feh}{\text{[Fe/H]}}
\newcommand{\msun}{M_\odot}
\newcommand{\sph}{$S_{\rm ph}$}

\newcommand{\Kepler}{\emph{Kepler}\xspace}
\newcommand{\Gaia}{\textit{Gaia}\xspace}

\shorttitle{Rotation of Solar Analogs}
\shortauthors{do Nascimento et al.}

\begin{document}
	
	\title{Rotation of solar analogs cross-matching Kepler and Gaia DR2}
	
	\correspondingauthor{J.-D.~do Nascimento, Jr.}
	\email{jdonascimento@cfa.harvard.edu}
	
	\author[0000-0001-7804-2145]{J.-D.~do~Nascimento,~Jr.}
	\affiliation{Universidade Federal do Rio Grande do Norte (UFRN), Departamento de F\'isica, 59078-970, Natal, RN, Brazil}
	\affiliation{Harvard-Smithsonian Center for Astrophysics, 60 Garden St., Cambridge, MA 02138, USA}
	
	\author[0000-0001-8179-1147]{L.~de~Almeida}
	\affiliation{Universidade Federal do Rio Grande do Norte (UFRN), Departamento de F\'isica, 59078-970, Natal, RN, Brazil}
	
	\author[0000-0002-9668-5547]{E.~N.~Velloso}
	\affiliation{Universidade Federal do Rio Grande do Norte (UFRN), Departamento de F\'isica, 59078-970, Natal, RN, Brazil}
	
	\author[0000-0003-1963-636X]{F.~Anthony}
	\affiliation{Universidade Federal do Rio Grande do Norte (UFRN), Departamento de F\'isica, 59078-970, Natal, RN, Brazil}
	
	\author[0000-0001-7152-5726]{S.~A.~Barnes}
	\affiliation{Leibniz Institute for Astrophysics Potsdam (AIP), Potsdam, Germany}
	
	\author{S.~H.~Saar}
	\affiliation{Harvard-Smithsonian Center for Astrophysics, 60 Garden St., Cambridge, MA 02138, USA}
	
	\author{S.~Meibom}
	\affiliation{Harvard-Smithsonian Center for Astrophysics, 60 Garden St., Cambridge, MA 02138, USA}
	
	\author[0000-0002-9830-0495]{J.~S.~da~Costa}
	\affiliation{Universidade Federal do Rio Grande do Norte (UFRN), Departamento de F\'isica, 59078-970, Natal, RN, Brazil}
	
	\author[0000-0002-1332-2477]{M.~Castro}
	\affiliation{Universidade Federal do Rio Grande do Norte (UFRN), Departamento de F\'isica, 59078-970, Natal, RN, Brazil}
	
	\author{J.~Y.~Galarza}
	\affiliation{Universidade de S\~ao Paulo, Instituto de Astronomia, Geof\'isica e Ci\^encias Atmosf\'ericas (IAG), Departamento de Astronomia, Rua do Mat\~ao 1226, Cidade Universit\'aria, 05508-900, SP, Brazil}
	
	\author{D.~Lorenzo-Oliveira}
	\affiliation{Universidade de S\~ao Paulo, Instituto de Astronomia, Geof\'isica e Ci\^encias Atmosf\'ericas (IAG), Departamento de Astronomia, Rua do Mat\~ao 1226, Cidade Universit\'aria, 05508-900, SP, Brazil}
	
	\author[0000-0003-4745-2242]{P.~G.~Beck}
	\affiliation{Institute of Physics, Karl-Franzens University of Graz, NAWI Graz, Universit\"atsplatz 5/II, 8010 Graz, Austria.}
	\affiliation{Instituto de Astrof\'{\i}sica de Canarias, E-38200 La Laguna, Tenerife, Spain}
	\affiliation{Departamento de Astrof\'{\i}sica, Universidad de La Laguna, E-38206 La Laguna, Tenerife, Spain}
	
	\author[0000-0002-4933-2239]{J.~Mel\'endez}
	\affiliation{Universidade de S\~ao Paulo, Instituto de Astronomia, Geof\'isica e Ci\^encias Atmosf\'ericas (IAG), Departamento de Astronomia, Rua do Mat\~ao 1226, Cidade Universit\'aria, 05508-900, SP, Brazil}
	
	\begin{abstract}
		A major obstacle to interpreting the rotation period distribution for main-sequence stars from Kepler mission data has been the lack of precise evolutionary status for these objects.
		We address this by investigating the evolutionary status based on Gaia Data Release 2 parallaxes and photometry for more than 30,000 Kepler stars with rotation period measurements.
		Many of these are subgiants, and should be excluded in future work on dwarfs.
		We particularly investigate a 193-star sample of solar analogs, and report newly-determined rotation periods for 125 of these.
		These include 54 stars from a prior sample, of which can confirm the periods for 50.
		The remainder are new, and 10 of them longer than solar rotation period, suggesting that sun-like stars continue to spin down on the main sequence past solar age.
		Our sample of solar analogs could potentially serve as a benchmark for future missions such as PLATO, and emphasizes the need for additional astrometric, photometric, and spectroscopic information before interpreting the stellar populations and results from time-series surveys.
	\end{abstract}
	
	
	\section{Introduction}\label{intro}
	
	\begin{figure*}[ht]
		\centering
		\includegraphics[width=\linewidth]{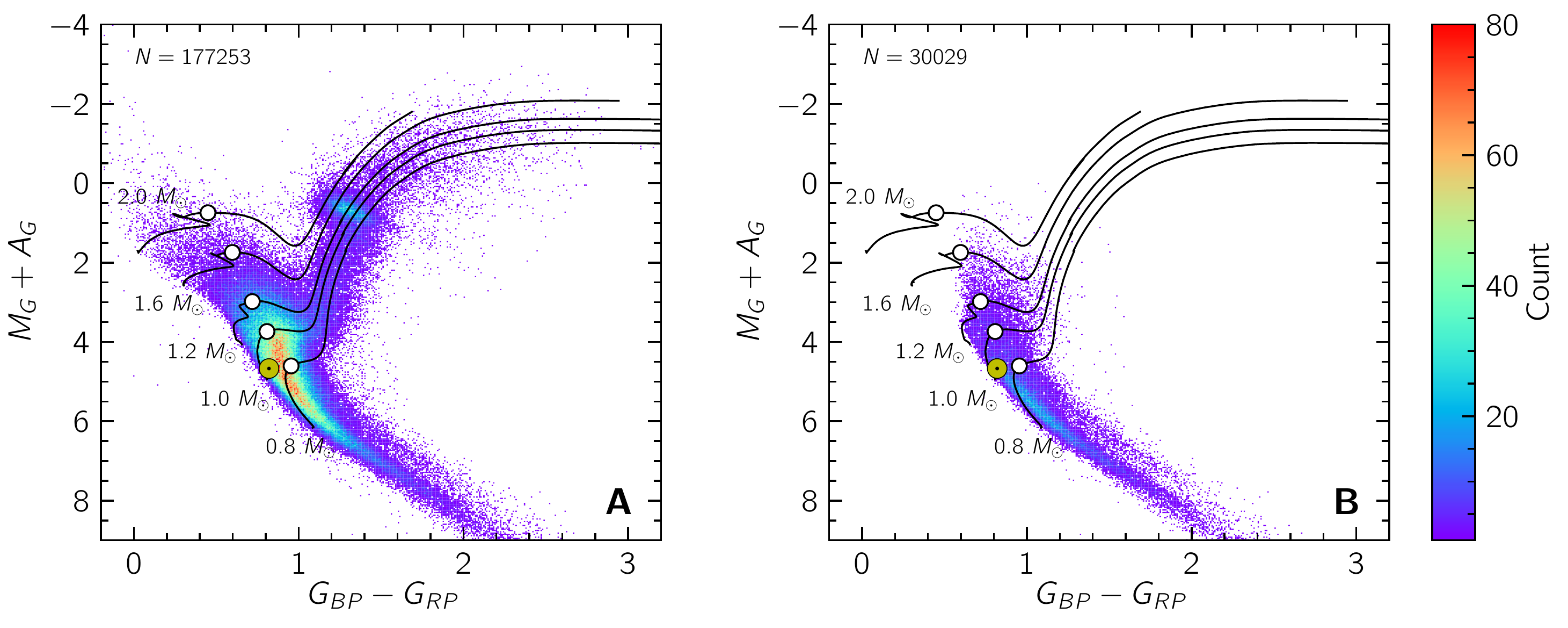}
		\caption{
			(a) Hertzsprung--Russell (HR) diagram for all matched \Kepler  objects in \Gaia DR2 (Sample 1).
			(b) HR diagram for the stars with rotation in MMA14. The white circles indicate where the subgiants region begins. The Sun is marked in yellow in both panels.
		}
		\label{fig:fig1}
	\end{figure*}
	
	Stellar rotation studies have experienced significant improvements thanks to new observational campaigns and theoretical developments.
	Earlier measurements of stellar rotation were typically performed by spectroscopy via  rotational broadening of absorption lines \citep[e.g.,][]{Struve1930, Carroll1933}.
	Even though it is very useful, this technique relies on prior knowledge of stellar radius and inclination and is limited to relatively fast rotators.
	
	In the last decade, with the advent of space-based photometric surveys, stellar rotation studies have advanced to a point where astronomers are able to determine rotational periods for a large number of stars simultaneously.
	This photometric technique is based on the stellar brightness modulation caused by non-uniformities on the surface due to transits of magnetic features (e.g. spots and plages).
	The repeated crossings of these active regions on the visible hemisphere as the star rotates induce a quasi-periodic variability in the photometric time series.
	The analysis of this modulation has been used to determine the stellar rotation period for an exceptionally large sample of stars, covering a wide distribution in masses, metallicities, and evolutionary states.
	Based on this technique, the CoRoT mission \citep{Baglin2006} was the first to contribute with studies of rotation periods for large ensembles of stars, and \cite{Affer2012} presented a catalog with 1,727 measurements of rotational periods for mid-F to mid-K stars.
	The \Kepler mission \citep{Borucki2010} has boosted studies of rotation due to its near-continuous observational cadence and coverage during four uninterrupted years.
	The studies of rotation in open clusters observed by \Kepler \citep{Meibom2011,Meibom2015,Barnes2016} were essential for empirical gyrochronology calibrations.
	Among the main large-scale rotation studies using \Kepler data are those of \cite{McQuillan2013,McQuillan2014} (henceforth MMA14), which provided rotation period measurements for 34,030 stars and discovered a bimodal period distribution for M dwarfs in the \Kepler field, which was subsequently confirmed to exist for K dwarfs as well.
	
	As underlined by \cite{Davenport2018}, this behavior has been suggested to be linked to a non-continuous age distribution for nearby stars, firstly by \cite{Hernandez2000} for nearby Hipparcos mission targets, or to a previously unknown phase of rapid angular momentum loss for low-mass stars, similar to the ``Vaughan-Preston'' gap seen in chromospheric activity indicators \citep{Vaughan1980}.
	\cite{Reinhold2019} found a transition from spot to faculae domination in solar-type stellar evolution, suggesting that intermediate rotation periods are largely undetected because dark spots and bright faculae cancel each other out at $\approx$800 Myr.
	These explanations are limited by the lack of precise age determinations for these stars.
	
	Many other key studies have also contributed with rotation periods for thousands of stars \citep[e.g.][etc]{Nielsen2013, Reinhold2013, Garcia2014, doNascimento2014, Aigrain2015}.
	Classical numerical algorithms have steadily improved for statistical detection and characterization of periodic signals in unevenly-sampled data.
	The Auto-Correlation Function \citep[ACF, e.g.][]{Brockwell2002} and generalized Lomb-Scargle periodograms \citep{Zechmeister2009} are widely employed to recover the rotational signal.
	The periods recovered from those different approaches generally agree very well for the most active stars.
	However, for less active and generally older objects, this operation is more challenging, because the ability to infer rotation depends on photometric variability and instrument precision.
	This raises a discussion about how common is the rotation of the Sun, a benchmark in stellar astrophysics research.
	
	Notably, old field stars appear to rotate more rapidly than predicted by the classical smooth spin-down laws.
	A transition in the rotational evolution after reaching a certain Rossby number (ratio of $\prot$ to convective turnover time) is apparently required in order to fit magnetic braking models to these stars \citep{vanSaders2016}.
	A reliable empirical calibration of gyrochronology is challenging and multiple relations are apparently required \citep{Angus2015}, hinting that the gyrochronology relationship is not complex enough.
	Recent studies of solar-type stars discuss a phase of stalled rotational evolution followed by an episode of rapid spin-down \citep{Metcalfe2019}.
	This transition seems to happen close to the age of the Sun.
	
	\begin{figure*}[ht]
		\centering
		\includegraphics[width=\linewidth]{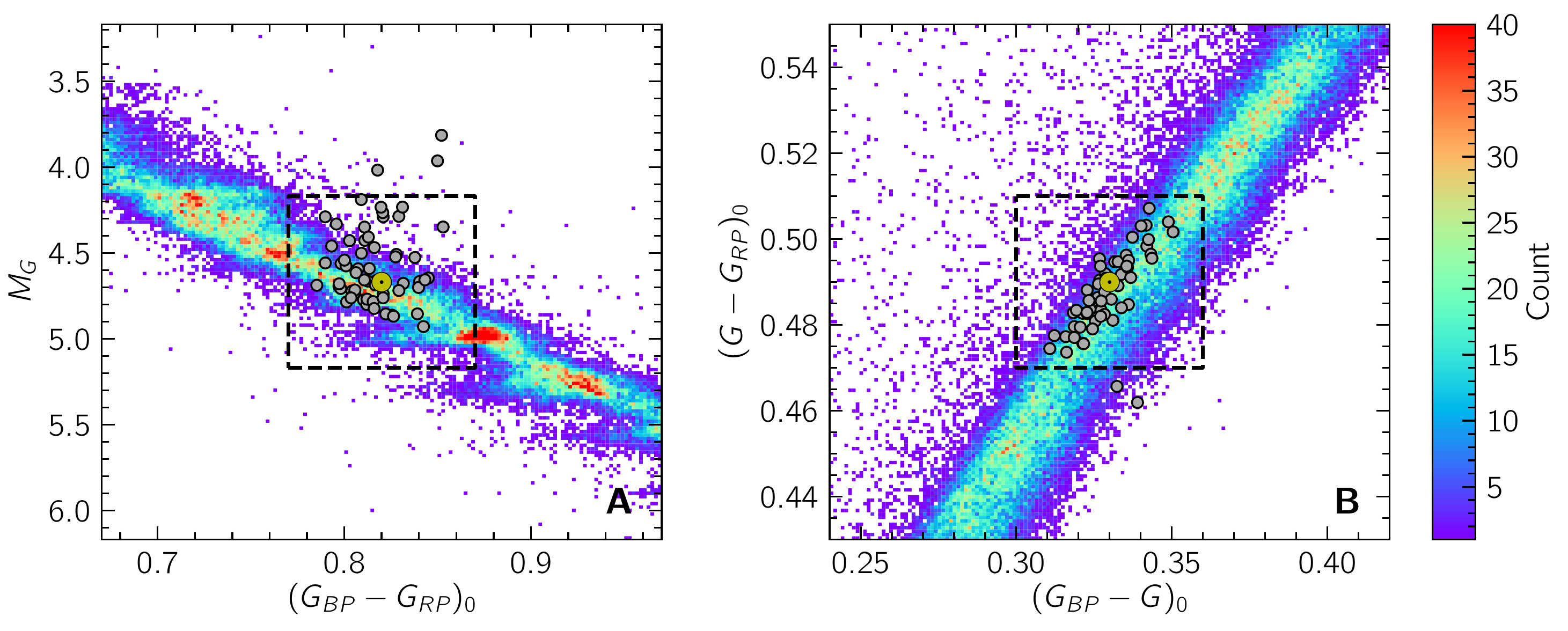}
		\caption{
			(a) Color-magnitude diagram and
			(b) color-color diagram showing the \Kepler--\Gaia cross-match sample near the solar region in each color from \Gaia DR2. The extinction $A_G$ and reddening $E(\mathrm{BP}-\mathrm{RP})$ were taken into consideration and only the stars for which these parameters were reported are shown. The subscript index 0 indicates reddening correction. The gray circles are the 61 solar twins from \cite{TucciMaia2016} which had \Gaia photometry and single status, and the dashed lines indicate where we performed our cut to select our sample of solar photometric analogs. The Sun is marked in yellow in both panels.
		}
		\label{fig:fig2}
	\end{figure*}
	
	In particular, solar twins allow us to decide to what extent the Sun itself can be considered a ``typical''  1.0~$\msun$ star \citep{Gustafsson2008}, and establishing a sample of solar analogs is important to map its past, present, and future \citep{Hardorp1978, Cayrel1996, Melendez2007, Monroe2013, doNascimento2013, doNascimento2014}.
	A sample of solar analogs, with determined $\prot$ similar to the one of the Sun, is  important to study the ``Sun in Time'' \citep[see][]{Dorren1994, PortodeMello1997, doNascimento2013, Beck2017}.
	
	In this context, the astrometric data from the \Gaia mission \citep{Gaia2016} help us to shed a new light on this stellar population puzzle.
	The \Gaia satellite observed over a billion stars in our Galaxy in order to better understand its structure, formation, and evolution.
	The \Gaia Data Release 2 \citep[hereafter DR2,][]{Gaia2018} contains 1.69 billion sources with positions and G-band photometry, 1.33 billion of which also have parallaxes and proper motions \citep{Lindegren2018}.
	These data improve our understanding of fundamental astrophysical parameters of stars observed by other observational programs via cross-matching, besides better clarifying the evolutionary status of stars and distinguishing main-sequence stars from evolved ones.
	In addition,  we can test the robustness of fundamental parameters from \Kepler stars, such as $\log(g)$, mass, and $\teff$ \citep{Creevey2013} and color-magnitude diagram models and color  calibrations \citep{Bertelli1999}.
	
	In this paper, we investigate the rotation of solar analogs and solar twin candidates cross-matching \Kepler and \Gaia and we carry out an analysis of their evolutionary states.
	Our goal is to contribute to the understanding of the rotational evolution along the main sequence and investigate how selection effects might have affected the distribution of rotation periods for solar analogs.
	
	\begin{figure*}[ht]
		\centering
		\includegraphics[width=0.7\linewidth]{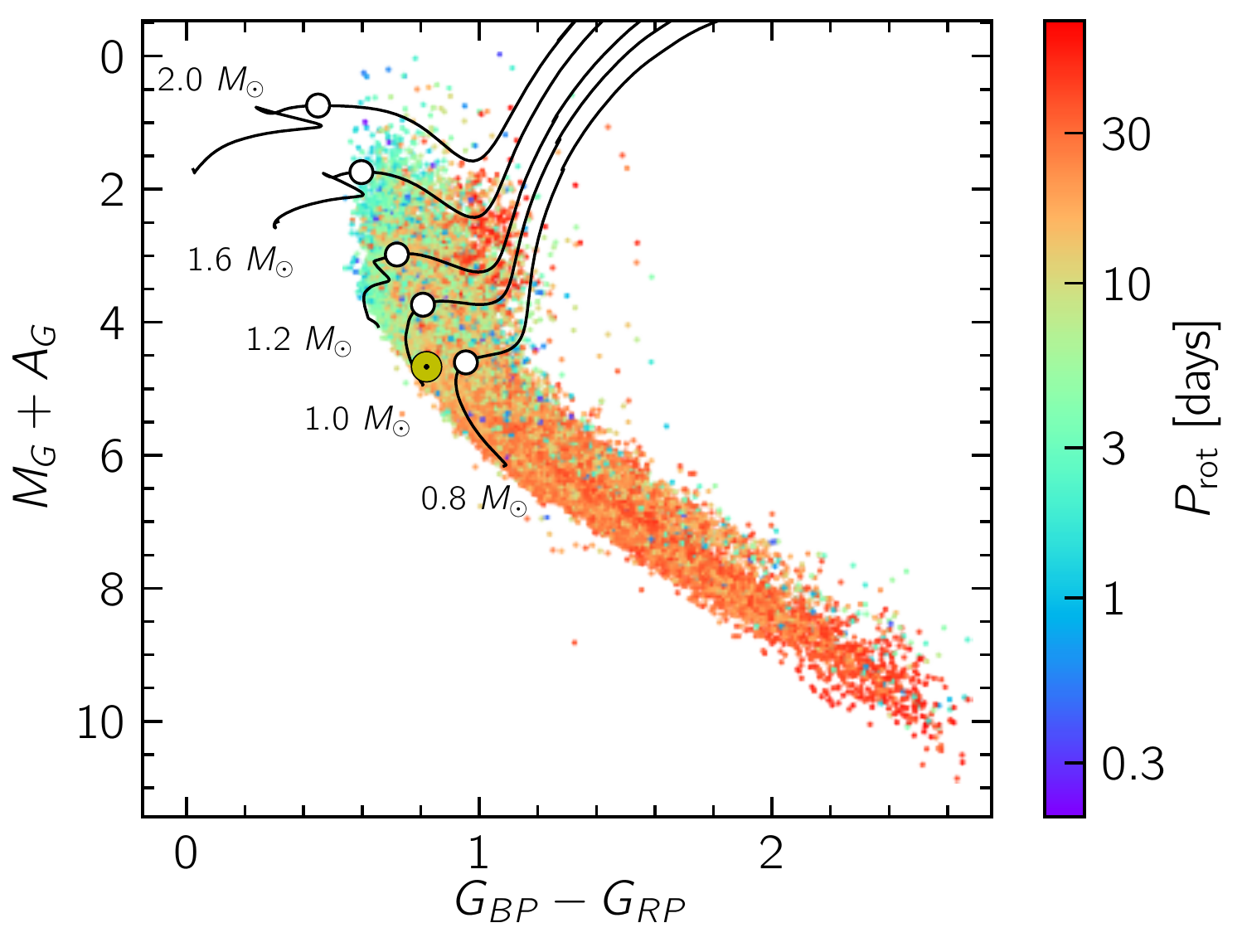}
		\caption{
			HR diagram using photometry from \Gaia DR2 and colored by their measured \Kepler rotation periods in MMA14. The solid black lines represent the evolutionary tracks for 0.8, 1.0, 1.2, 1.6, and 2.0 $\msun$ respectively.
			The white circles indicate where the subgiants region begins.
		}
		\label{fig:fig3}
	\end{figure*}
	
	\section{Observational Data}
	\label{obsv}

	\subsection{Sample selection}
	\label{sec:data}
	
	We cross-matched the positions of stars from the NASA Exoplanet Archive catalog, consisting of all stellar targets with long-cadence observations from the original \Kepler field, to the \Gaia DR2 within a search radius of 1 arcsecond.
	The targets with multiple matches were selected by minimizing the angular distance between the \Kepler and the \Gaia objects.
	Of the 197,775 stars found in this cross-match, we removed the ones that had negative or poorly determined parallaxes ($\varpi/\sigma_\varpi < 5$).
	In order to obtain parameters which are compatible with \Gaia photometry while still taking advantage of the vast amount of photometric, spectroscopic and asteroseismic analyses performed within the \Kepler field in the last decade, we decided to use the revised temperature values from \cite{Berger2018}. This left us with a total of 177,253 stars, which constituted the \Kepler sample we adopted.
	From now on we will refer to this sample as Sample 1.
	As was mentioned before, MMA14 determined rotation periods for 34,030 stars from \Kepler, of which 30,029 ($\approx$88\%) were present in our Sample 1.
	The Hertzsprung-Russell (HR) diagrams for Sample 1 and for the ones recovered from MMA14 are shown in Figure \ref{fig:fig1}.
	
	We then proceeded to select a subset of objects similar enough to our Sun to be considered solar analogue candidates.
	To this end, we found a match to 61 single twins studied by \cite{TucciMaia2016} in \Gaia DR2 and we used them after applying the zero-point corrections from \cite{Weiler2018} in order to calibrate a photometric standard for our classification (see Figure \ref{fig:fig2}).
	\cite{Andrae2018} suggested a slightly different set of criteria which did not encompass some sun-like stars, and for this reason we decided to apply our twin-adjusted criteria specified in Table \ref{tab:sample}.
	The solar photometry in the \Gaia pass-bands was obtained by \cite{Casagrande2018}.
	In Figure \ref{fig:fig2} we plot the stars from Sample 1 for which extinction $A_G$ and reddening $E(\textrm{BP}-\textrm{RP})$ were reported in that data release.
	We also calculated the reddening correction for the two colors $(G_{\rm BP}-G)$ and $(G-G_{\rm RP})$ from the extinction coefficients as functions of $\teff$ and $\feh$ as in \cite{Casagrande2018}.
	A subset of 3557 objects was then selected by applying the criteria from Table \ref{tab:sample} to select solar analogs from Sample 1.
	Those were further restricted to a set of 193 stars with small temperature uncertainties ($\sigma_{\teff} \leq 150$ K) and which had well-determined ($\sigma_{\feh} \leq 0.15$ dex) spectroscopic metallicities from the \Kepler Stellar Properties Catalogue \citep[KSPC DR25,][]{Mathur2017}, so that we could perform a precise evolutionary analysis ({described in} Section \ref{sec:end} below).
	This second and last sample will be referred to as Sample 2.
	Table \ref{tab:sample} details the source count in every step of our selection.
	
	\begin{deluxetable*}{@{}l@{}r}
		\tablecaption{Number of stars in each filter step and criteria for analogue candidates. \label{tab:sample}}
		\tablehead{
			\colhead{Filter} & \colhead{Source Count}
		}
		\startdata
		\Gaia Data Release 2 (GDR2) \dotfill & \dotfill 1,692,919,135 \\
		GDR2 with five-parameter astrometric solution \dotfill & \dotfill 1,331,909,727 \\
		\Kepler stars cross-matched within 1 arcsec \dotfill & \dotfill 197,775 \\
		Restraining to $\varpi/\sigma_\varpi > 5$ \dotfill & \dotfill 187,497 \\
		Temperature values in \cite{Berger2018} (Sample 1) \dotfill & \dotfill 177,253 \\
		Dereddened photometric solar analogs (see below) \dotfill & \dotfill 3,557 \\
		$\sigma_{\teff}\leq 150$ K  and $\sigma_{\feh}\leq0.15$ dex (Sample 2) \dotfill & \dotfill 193 \\
		\midrule
		\hspace{0.2cm} Photometric Solar Analogue Candidates Criteria & Solar Values \hspace{0.2cm} \\
		$|\teff - 5772| < 150$ & $T_{\rm eff\odot}=5772$ K\\
		$0.77 < (G_{\rm BP} - G_{\rm RP})_0 < 0.87$ &  $(G_{\rm BP} - G_{\rm RP})_\odot=0.82$\\
		$4.17 < M_G < 5.17$  & $M_{G\odot}=4.67$\\
		$0.30 < (G_{\rm BP} - G)_0 < 0.36$ & $(G_{\rm BP}-G)_\odot=0.33$\\
		$0.47 < (G - G_{\rm RP})_0 < 0.51$ & $(G-G_{\rm RP})_\odot=0.49$
		\enddata
	\end{deluxetable*}
	
	\subsection{Evolutionary status of the sample}
	\label{Evolutionary_states}
	
	To determine the evolutionary status of our sample, we used evolutionary track models from \cite{Girardi2000}.
	They include an initial composition with $Z=0.019$ and $Y=0.273$.
	
	We used models with stellar masses ranging from 0.8 to 2.0 $\msun$ and solar metallicity, encompassing most of the stars contained in the present working sample and shown in the HR diagrams in Figure \ref{fig:fig1}.
	The turn-off point separating main-sequence and subgiant stars is defined as the age when the hydrogen content equals zero at the center of the model.
	From Figure \ref{fig:fig1}b, we can see that most stars are located below the turnoff line and therefore appear to be genuine dwarfs.
	On the other hand, stars located on the right-hand side of the  turn-off  indicate a significant contamination of this sample by stars on the subgiant branch \citep{Davenport2018, vanSaders2019}.
	
	\subsection{Rotation of Solar Analogs, Solar Twins, and Subgiants}
	\label{sec:types}
	
	Using \Gaia photometry and parallaxes in Figure~\ref{fig:fig3} we show for the first time the $\prot$ distribution within an HR diagram for stars in MMA14, superposed with our sample of new solar analogue candidates.
	Even though MMA14 attempted to choose their sample to be composed mainly by dwarfs at the main sequence, we can clearly see that most stars with  $\prot>30~\textrm{days}$, the red dots in Figure~\ref{fig:fig3}, are in fact subgiants or cool low mass M dwarfs.
	This substantial contamination by subgiants found here agree well with previous studies from \cite{Ciardi2011}, \cite{Mann2012} and \cite{Davenport2018} and has consequences on further studies based on MMA14 \citep[e.g.][]{vanSaders2016}.
	The binary sequence is also clearly visible.
	Furthermore, another important aspect to be discussed is the lack of solar analogs with rotation periods comparable to solar.
	Theoretical models take into account the mass scaling of stellar wind torque and the braking dependence on Rossby number in order to explain this observed rotation distribution \citep{Barnes2010,Matt2015}.
	This point has implications for age-dating of field stars \citep[e.g.][]{vanSaders2016, vanSaders2019}.
	Because of this, in this study we attempted to carefully determine the $\prot$ of the 193 photometric solar analogue candidates which constituted our Sample 2 (see Section \ref{sec:data}).
	
	\begin{figure*}[ht]
		\centering
		\includegraphics[width=\linewidth]{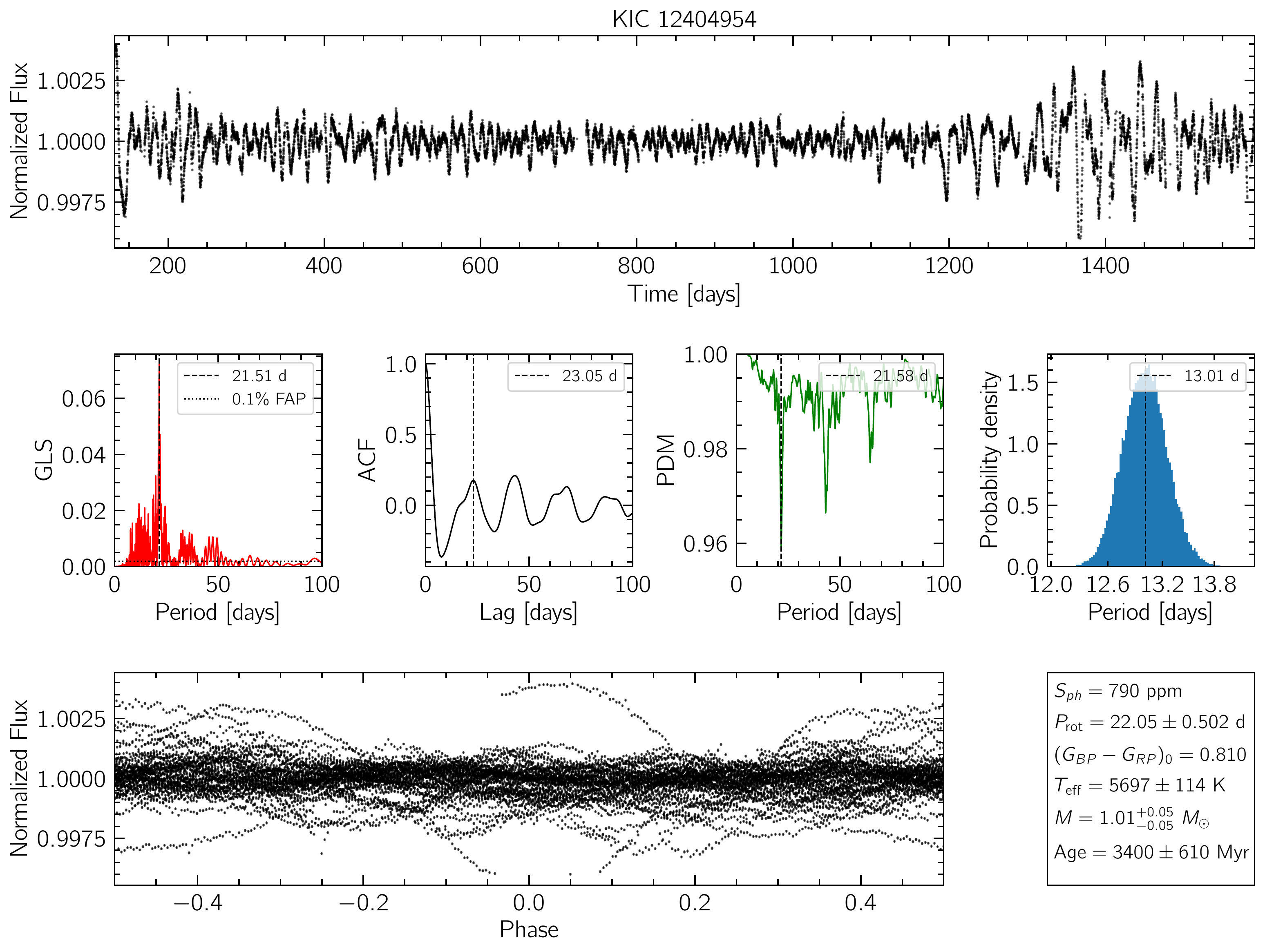}
		\caption{
			Analysis of sample light curve of a star from Sample 2 (KIC 12404954). \emph{Top}---All available long cadence data for this target.
			\emph{Middle}---Four different methods of $\prot$ determination are illustrated for this star: the Auto-Correlation Function smoothed by a Gaussian kernel, the generalized Lomb-Scargle periodogram  with False Alarm Level 0.1\%, the $\Theta$ statistic from PDM, and the posterior probability distribution for the Gaussian Process regression.
			\emph{Bottom left}---Phase-folded light curve.
			\emph{Bottom right}---Reference information for this star.
			Versions of this figure panel for every target analyzed (193 images) are available as an electronic figure set
			in the online Journal.
		}
		\label{fig:fig4}
	\end{figure*}
	
	\section{Rotation period measurements}
	\label{sec:rot}
	
	The study from MMA14 performed an automatic ACF analysis of 34,030 \Kepler stars cooler than 6500 K.
	However, this method has several deficiencies, mainly the need for several heuristic choices such as how to smooth the ACF and how to define and select a peak.
	
	On the other hand, the generalized Lomb-Scargle periodogram avoids these limitations, besides providing a simple way of assessing period uncertainties based on peak widths.
	This comes with the drawback of assuming a perfectly sinusoidal signal, which is not necessarily the case, and the spot dynamics can bring up higher peaks in alias frequencies and harmonics.
	
	Another obstacle when processing photometric time series is the intrinsic noise and systematic errors introduced by the \Kepler spacecraft and its quarterly CCD change for a given target.
	The latest presearch data conditioning algorithm, multiscale MAP \citep[msMAP,][]{Stumpe2014} is widely employed to correct and detrend \Kepler light curves, and, although it is optimized to search for exoplanetary transits, it has been shown to preserve rotational signals to a certain degree.
	
	In our analysis, we applied both generalized Lomb-Scargle periodograms and the ACF to Sample 2.
	We used msMAP corrected light curves, stitching each quarter with their mean flux normalized to 1.0.
	No pre-processing was performed except for the clipping of outliers and the binning of the light curve to $\approx$2 hour cadence.
	We calculate the $\Theta$ statistic as defined in \cite{Stellingwerf1978} in order to check if the best period also optimizes phase dispersion.
	In Figure \ref{fig:fig4} we show a characteristic example of a light curve from our sample, together with the periodogram, autocorrelation and phase dispersion minimization (PDM) analysis.
	
	We also demonstrate a Gaussian Process regression technique as in \cite{Angus2018} using the quasi-periodic sum of exponentials kernel from \cite{Foreman-Mackey2017}, which can be implemented with linear complexity using optimizations found in the \texttt{celerite} package.
	We used an MCMC sampling of the posterior distribution with 32 walkers, 5000 steps, and 500 burn-in samples.
	It is worth noticing that, as a distribution over $\ln{\prot}$, the uncertainties get larger for slower rotating stars.
	
	Each of these four methods explores different features of a periodic signal.
	This means that, if each one of them gives diverging answers for the same signal, even if all of them provide ``decent'' estimates within some individual metric, the periodic content of the signal should be understood to be either multiple or none at all.
	With that in mind, after determining the four period estimates, if the periodogram's False Alarm Probability (FAP) is below 0.1\%, we test for convergence by checking whether or not there is a single value to which at least three estimates approach.
	If there are two distinct sets converging to two different values, we report the one containing the ACF estimate, as it generally deals best with frequency harmonics.
	Out of the 193 analogue candidates, we report periodicities for 125 of them ($\approx$65\%).
	
	For each light curve we calculated a measure of photometric variability as a proxy of stellar activity, defined as the standard deviation of the light curve corrected by subtracting the photon noise \citep[\sph, see e.g.][]{Mathur2014}.
	Also, for the ones with determined periods, we derive an approximate age from the empirical gyrochronology formulation of \cite{Barnes2010}. 
	In the bottom right panel of Figure \ref{fig:fig4} these parameters are presented as reference information for each star.
	
	\begin{figure}
		\centering
		\gridline{
			\fig{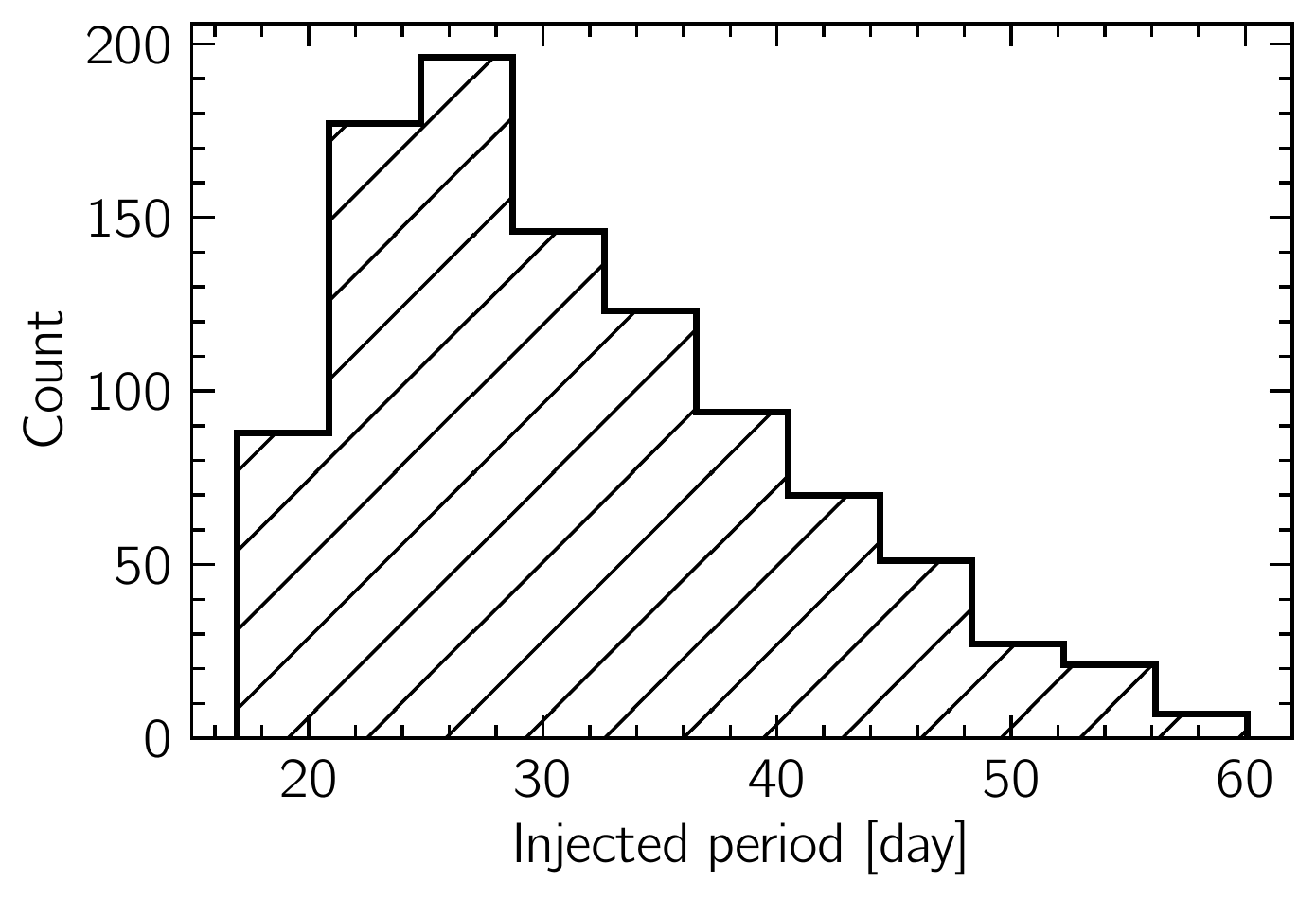}{0.4\textwidth}{(a) Histogram with the distribution of rotation periods from simulated light curves generated to validate our method.}
		}
		\gridline{
			\fig{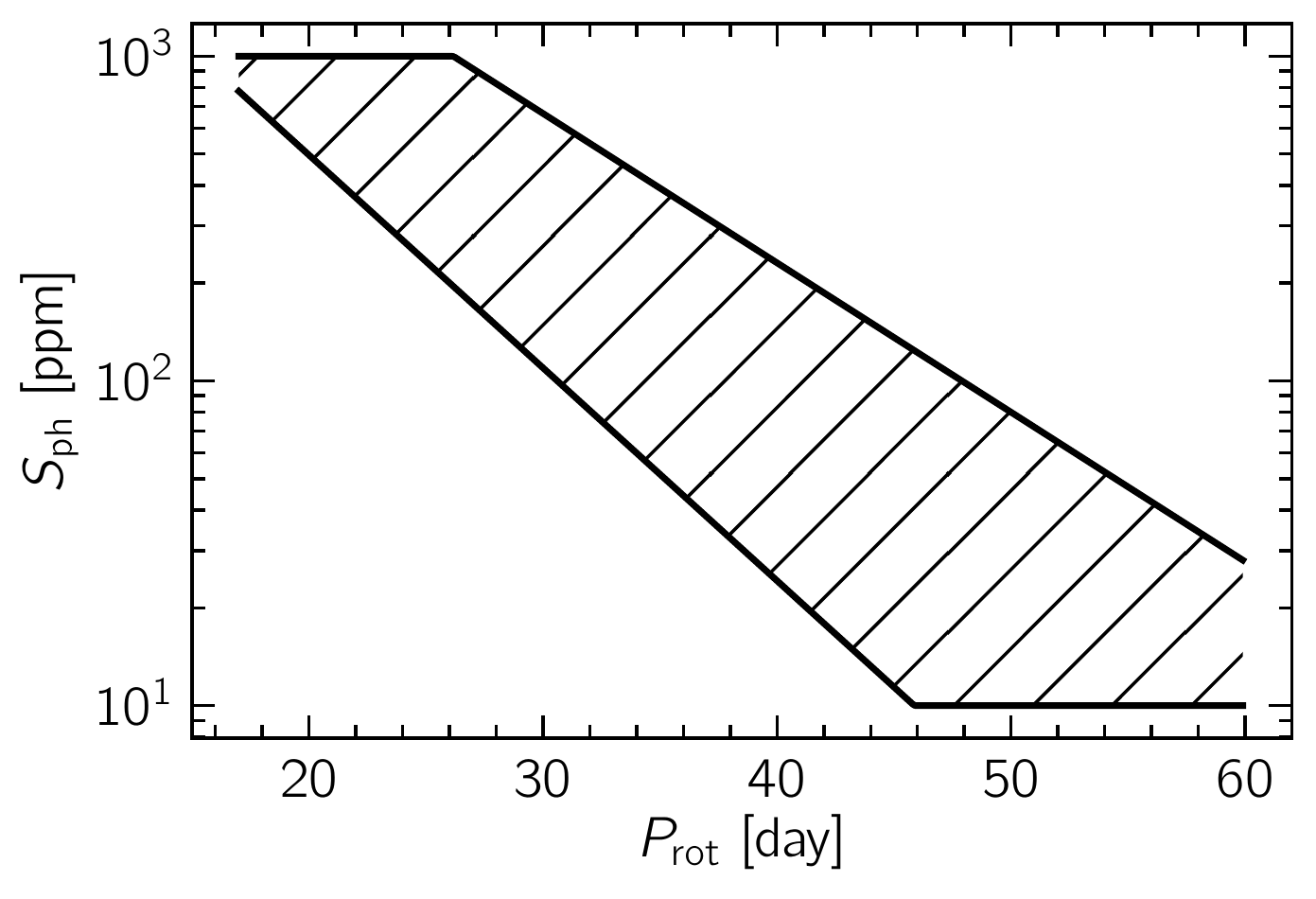}{0.4\textwidth}{(b) Region of activity-rotation diagram from which we sampled.}
		}
		\gridline{
			\fig{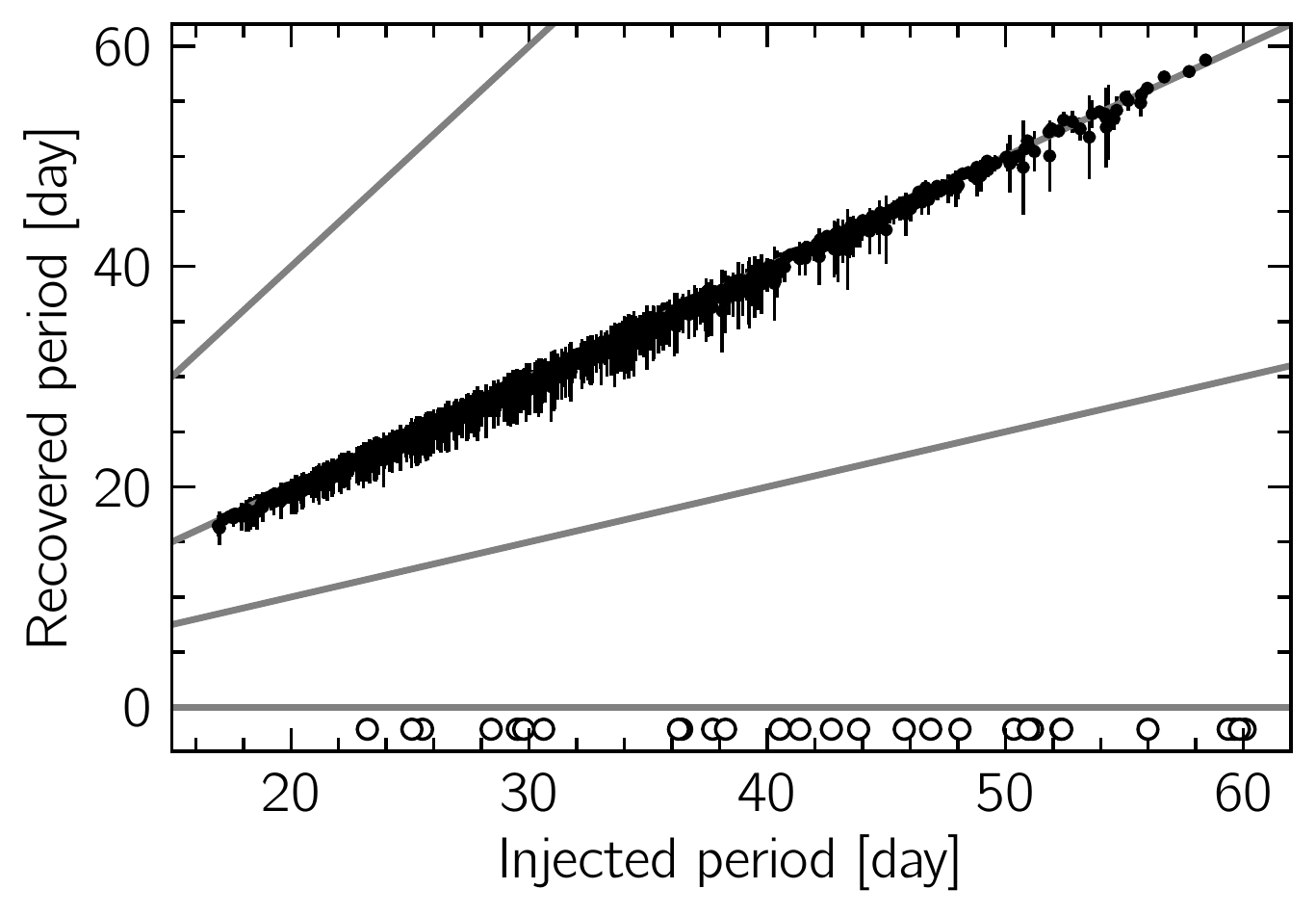}{0.4\textwidth}{(c) Recovered periods vs injected periods. The gray lines help to visualize the true period as well as the double- and half-frequency harmonics. The white circles indicate failed detections.}
		}
		\caption{}
		\label{fig:injec}
	\end{figure}
	
	To test the detectability of stellar rotation periods from simulated light curves, we present here the results of a signal injection and recovery exercise.
	We generated 1000 light curves based on the observed rotation period distribution and the corresponding expected photometric variability. Models using four spots were used and \Kepler-like noise was added.
	Figure \ref{fig:injec}a shows the distribution of periods generated using our simulator, ranging from 17 to 60 days, and Figure \ref{fig:injec}b indicates the region of the activity-rotation diagram from which the parameters were sampled.
	The recovered periods are compared with the injected ones in Figure \ref{fig:injec}c, where for only 29 cases was no period detected.
	No errors were incurred in terms of frequency harmonics and there appears to be no preferential region where the detection fails.
	
	\begin{figure}[ht]
		\centering
		\plotone{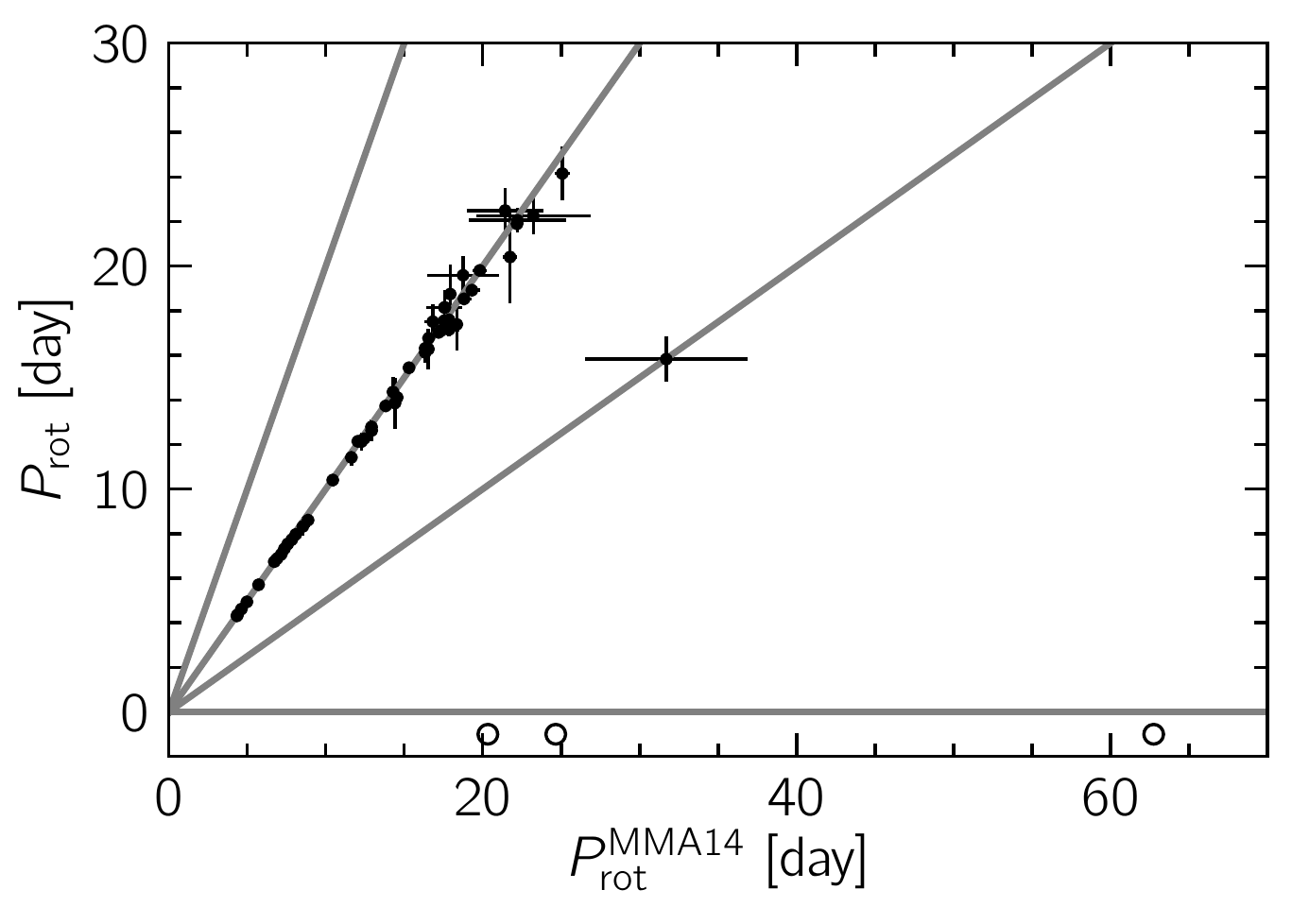}
		\caption{Comparison between the $\prot$ values determined using our methods and the ones reported in MMA14 for 54 analogs.}
		\label{fig:comparison}
	\end{figure}
	
	For further validation of our methodology, we compare our results to the ones reported by MMA14 for the 54 stars which were common between our samples. This comparison is illustrated in Figure \ref{fig:comparison}, where the gray lines mark 1:1, 1:2 and 2:1 relationships.
	As expected, we agree on the same value for 50 of those, and find exactly the second harmonic for KIC~4759349. There are also three cases in which we cannot confidently report a period due to diverging results, represented by the white circles.
	
	Since most active stars rotate with short periods, we claim that the methodology from MMA14 favors the detection of rotation periods on more active, young objects.
	As mentioned in their paper, many additional periods could have been detected but were missed due to the automatic nature of their procedures, predominantly long-period or low-amplitude signals.
	In addition, rotation periods determined by MMA14 requires the signal to repeat itself across several quarters of the \Kepler light curves, whereas rotation for old stars like the Sun has its modulation dependent on the phase of the magnetic cycle.
	Lastly, by using the old PDC-MAP pipeline \citep{Smith2012} and only three years of data from \Kepler, slower rotations inevitably become more challenging to detect. This also means, however, that even our new results cannot completely describe this tail of the distribution, given that even the latest correction algorithms work as high-pass filters.

	\begin{figure}
		\centering
		\includegraphics[width=\linewidth]{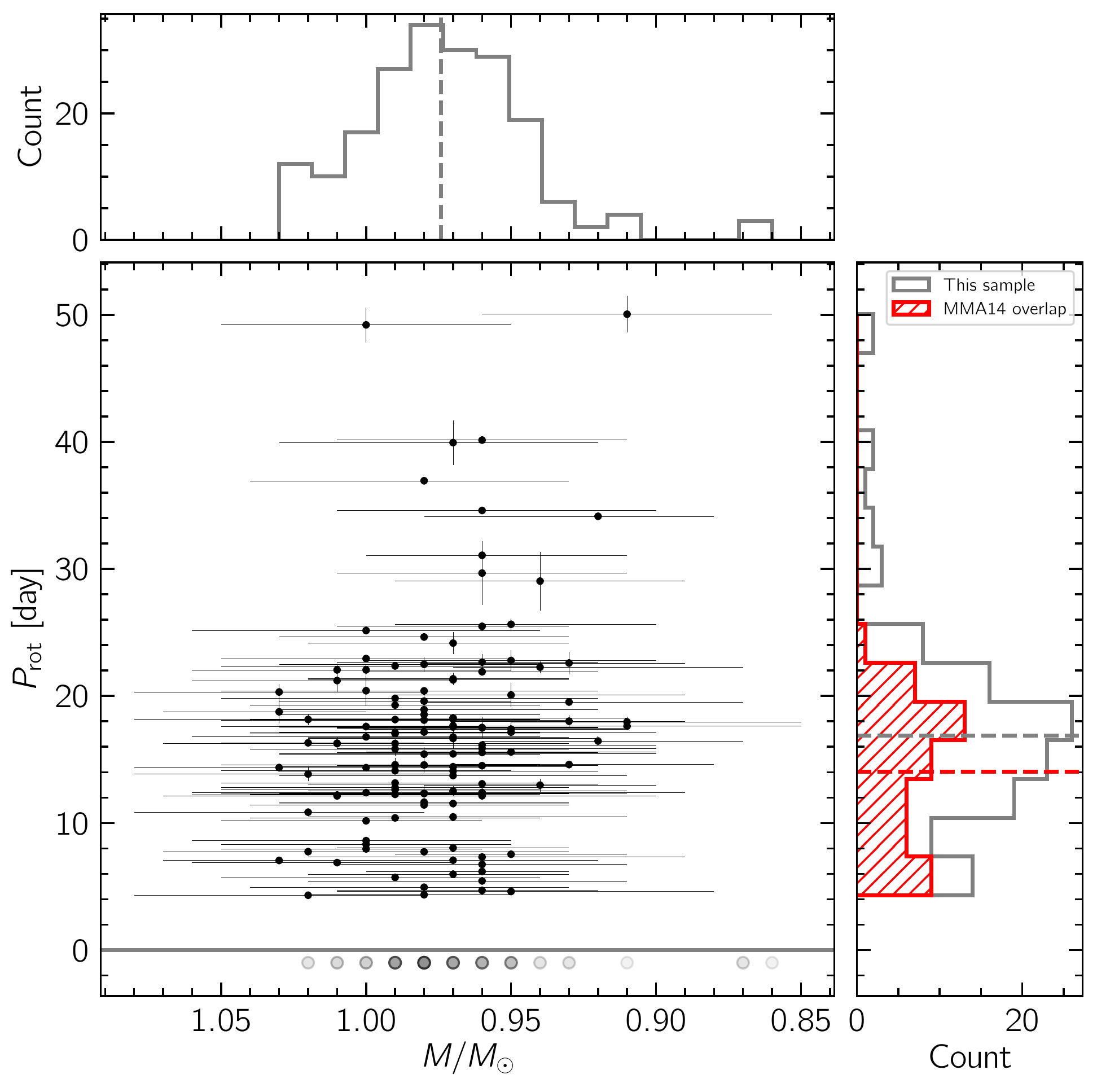}
		\caption{Rotation period distribution vs mass for our sample of 193 analog stars. The gray circles represent the ones without period detections. Both marginal distributions are projected to the gray histograms together with their averages (dashed lines). The masses average to $\langle M \rangle = 0.97\msun$. The average of the 125 rotation periods in this sample is $\langle\prot\rangle=16.87$ days, and the 51 periods which overlap with MMA14 are shown in red with their average of 14.02 days.}
		\label{fig:fig7}
	\end{figure}

	\begin{figure}
		\centering
		\includegraphics[width=\linewidth]{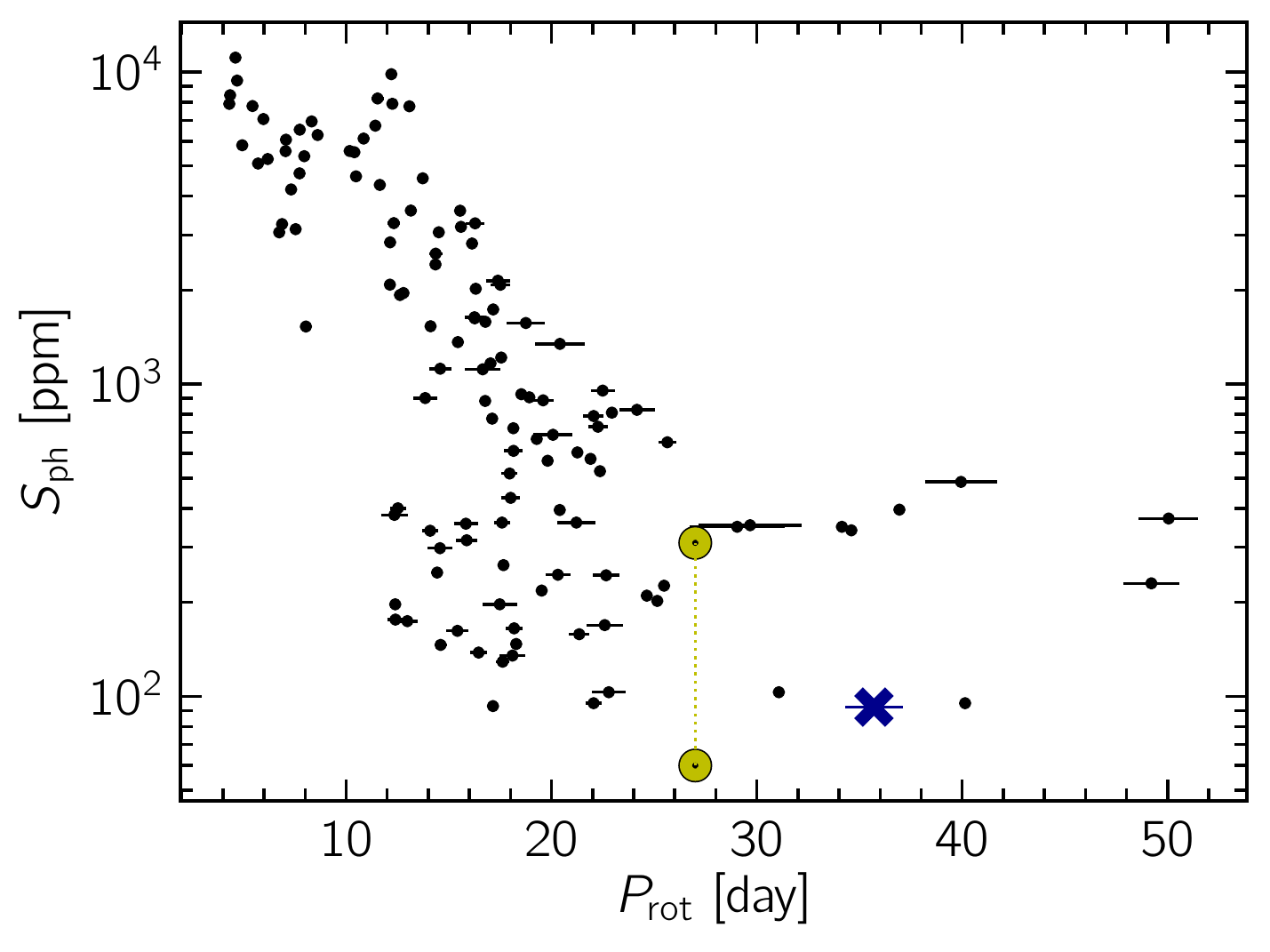}
		\caption{Photometric activity proxy \sph\ and its relationship with rotation for 125 analogs. The minimum and maximum activity of the Sun are marked in yellow for reference, as well as the blue cross corresponding to 8~Gyr-old solar twin HIP~102152.}
		\label{fig:fig8}
	\end{figure}
	
	\section{Results}
	\label{sec:end}
	
	\begin{deluxetable*}{cCCCccccccccc}
		\tablecaption{Derived masses, rotation periods, and summary of astrophysical parameters for 193 candidate solar analogs. \label{tab:results}}
		\tablehead{
			\colhead{KIC ID} & \colhead{Mass} & \colhead{$\teff$} & \colhead{$\feh$} & \colhead{$(G_{\rm BP} - G_{\rm RP})_0$} & \colhead{FAP} & \colhead{\sph} & \colhead{$\prot^\text{GLS}$} & \colhead{$\prot^\text{ACF}$} & \colhead{$\prot^\text{PDM}$} & \colhead{$\prot^\text{GP}$} & \colhead{$\prot$} & \colhead{Age$_\text{B10}$} \\
			\colhead{} & \colhead{($\msun$)} & \colhead{(K)} & \colhead{} & \colhead{} & \colhead{(\%)} & \colhead{(ppm)} & \colhead{(d)} & \colhead{(d)} & \colhead{(d)} & \colhead{(d)} & \colhead{(d)} & \colhead{(Myr)}
		}
		\startdata
		1434277 & 0.99^{+0.05}_{-0.05} & 5752\pm115 & \phantom{-}0.14\pm0.15 & 0.821 & 0.0 & 5530 & 10.65 & 10.46 & 10.50 & 10.00 & $10.40\pm0.14$ & $790\pm160$\\
		1867120 & 0.98^{+0.06}_{-0.04} & 5830\pm117 & -0.54\pm0.15 & 0.794 & 0.1 &   90 & 46.94 & 49.53 & 47.03 & --- & --- & --- \\
		1872084 & 0.97^{+0.06}_{-0.07} & 5706\pm114 & \phantom{-}0.07\pm0.15 & 0.805 & 0.0 &  464 & 14.03 & 25.26 & 50.58 & --- & --- & --- \\
		2161400 & 0.92^{+0.04}_{-0.05} & 5648\pm113 & -0.36\pm0.15 & 0.820 & 0.0 &  138 & 16.99 & 16.59 & 16.95 & 15.27 & $16.45\pm0.40$ & $1500\pm230$
		\enddata
		\tablecomments{This table is available in its entirety in a machine-readable form in the online jounal. A portion is shown here for guidance regarding its form and content.}
	\end{deluxetable*}
	
	We report new rotation period determinations for 125 solar analogs, 74 of which were not previously analyzed in terms of their rotation and evolutionary status based on an HR diagram.
	These determinations and related characteristics of the stars are plotted in Figures \ref{fig:fig7} and \ref{fig:fig8}.
	In Table \ref{tab:results} we detail the star identifiers and their main astrophysical quantities together with our $\prot$ determinations and uncertainties using each method.
	For these stars we also present the masses obtained from evolutionary tracks due to \cite{Kim2002} and \cite{Yi2003}, following the same procedure described in \cite{Grieves2018}.
	Although the statistical mass distribution is slightly skewed towards lower masses due to the non-linear magnitude-mass relationship ($\langle M \rangle = 0.97\msun$; see Figure \ref{fig:fig7}), virtually all of them ($\approx$98\%) lie within $0.90 < M/\msun < 1.10$.
	The typical mass error in this analysis is $\approx$0.06 $M_\odot$.
	
	Whereas \cite{Reinhold2013}, \cite{Nielsen2013} and \cite{McQuillan2013} have claimed the absence of slow rotation periods for solar analogs from \Kepler, this paper shows that there are many stars on the main-sequence that are spinning slower than the Sun.
	That seems to suggest that stars continue to slow down on the main-sequence after solar age, even if they are hard to detect, in agreement with \cite{Lorenzo-Oliveira2019}.
	The average rotation value is significantly different from the ones common to MMA14, presented by a red line in Figure~\ref{fig:fig7}.
	We verified that there is no bias in mass between the two samples that would account for this difference between the mean periods.
	
	It is worth noticing that this method of identifying solar analogs relies deeply on the precision of the photometry, and passband calibrations \citep[e.g.][]{Weiler2018} together with precise reddening corrections play an essential role.
	This is also true for any subsequent theoretical modeling and evolutionary analysis. We hence propose this set of stars as a benchmark catalog to contribute in future careful observations to twin-based calibrations.
	
	Lastly, Figure~\ref{fig:fig8} presents the measured photometric activity proxy and its relationship with the reported rotation periods.
	The value of \sph\ for the Sun is derived from the VIRGO light curve and varies approximately from 60 to 310 ppm during its activity cycle.
	For comparison purposes, we have included the star HIP~102152 studied in \cite{Lorenzo-Oliveira2020} as a blue cross in Figure \ref{fig:fig8}. This 8 Gyr-old solar twin represents a much older Sun. Its rotation was determined from spectroscopic activity modulations (Ca \textsc{ii} H \& K) and we derive the photometric \sph\ from TESS light curves.
	Given that these stars are all photometric analogs, i.e., have roughly the same color, temperature, and mass as the Sun, the variations observed in rotation and activity correspond to age variations.
	These follow the expected decline in activity and corresponding braking of the rotation with age, with a detectability threshold around the activity level of the Sun.
	This relates to the scarcity of stars rotating slower than the Sun in most $\prot$ surveys.
	
	\section{Conclusions}
	
	This benchmark of solar analogs will be useful in further investigations of the effects of metal content in stellar rotation and activity.
	In addition, the rotation period distribution appears to imply that there is a detectability threshold around the variability level of the Sun which biases current interpretations of empirical gyrochronology.
	Thus, this sample should also be a priority in investigations on transitions in the rotational evolution of the Sun.
	Further observations can provide us with better precision in fundamental parameters (spectroscopically) and the instrumental effects in long period light curves (photometrically, e.g. with TESS and PLATO).
	
	\acknowledgements
	Thanks to Megan Bedell for her helpful discussions and for the \url{gaia-kepler.fun} cross-match database.
	JDN is supported by a CNPq PQ1 physics and astronomy Fellowship.
	SHS is grateful for support from NASA Heliophysics LWS grant NNX16AB79G.
	This work has made use of data from the European Space Agency (ESA) mission \Gaia (\url{http://www.cosmos.esa.int/gaia}), processed by the \Gaia Data Processing and Analysis Consortium (DPAC, \url{http://www.cosmos.esa.int/web/gaia/dpac/consortium}).
	Funding for the DPAC has been provided by national institutions, in particular the institutions participating in the \Gaia Multilateral Agreement.

\end{document}